\documentclass[fleqn,10pt]{wlscirep}
\usepackage{graphicx}
\usepackage{epsfig}
\usepackage{epstopdf}
\usepackage{amsmath}
\usepackage{pdfpages}

\title{Doping-induced perturbation and percolation in the two-dimensional Anderson lattice}

\author[1,2]{Lan-ying Wei}
\author[1,2,3,*]{Yi-feng Yang}
\affil[1]{Beijing National Laboratory for Condensed Matter Physics and Institute of Physics, Chinese Academy of Sciences, Beijing 100190, China}
\affil[2]{School of Physical Sciences, University of Chinese Academy of Sciences, Beijing 100190, China}
\affil[3]{Collaborative Innovation Center of Quantum Matter, Beijing 100190, China}

\affil[*]{yifeng@iphy.ac.cn}

\begin{abstract}
We examine the doping effects in the two-dimensional periodic Anderson model using the determinant Quantum Monte Carlo (DQMC) method. We observe bound states around the Kondo hole site and find that the heavy electron states are destroyed at the nearest-neighbor sites. Our results show no clear sign of hybridization oscillation predicted in previous mean-field calculations. We further study the electron transport with increasing doping and as a function of temperature and obtain a critical doping $x_c\approx 0.6$ that marks a transition from the Kondo insulator regime to the single-ion Kondo regime. The value of $x_c$ is in good agreement with the predicted threshold for the site percolation. Our results confirm the percolative nature of the insulator-metal transition widely observed in doped Kondo insulators.
\end{abstract}
\begin{document}

\flushbottom
\maketitle

\thispagestyle{empty}

\section*{Introduction}
Chemical substitution or doping of the magnetic ions (Ce, Yb, U, ...) by its nonmagnetic counterpart element (La, Th, Y, ...) gives rise to local vacancies of the magnetic $f$-moments (called the Kondo holes) and could cause dramatic changes in the ground state properties of heavy fermion compounds. In Kondo insulators such as CeNiSn \cite{echizen1999effect, adroja1996comparative, goraus2013quantum}, CeRhSb \cite{adroja1996comparative, goraus2013quantum}, and Ce$_{3}$Bi$_{4}$Pt$_{3}$ \cite{pietrus2008kondo}, interesting new physics has been proposed such as the bound states near the Kondo holes \cite{sollie1991simple,sollie1991local,schlottmann1992impurity,schlottmann1996influence,schlottmann1996metal}. With increasing concentrations of doping, an insulator-to-metal transition has been widely observed, accompanying with the change from the dense Kondo lattice regime to the single-ion Kondo regime. This transition has been generally speculated to be of percolative nature \cite{aharony2003introduction}. Many work, mostly based on mean-field calculations, have been carried out in order to clarify the relevant physics \cite{schlottmann1992impurity,li1994alloying}. To the best of our knowledge, no exact numerical calculations has been done. In particular, it is not clear how Kondo holes may destroy the many-body heavy electron state and, with increasing doping, drive the system from the Kondo lattice physics to the single-ion Kondo physics.

In this work, we use the determinant Quantum Monte Carlo (DQMC) method to study the electronic and transport properties of the doped two-dimensional periodic Anderson lattice. DQMC is an exact numerical method for a finite lattice but often suffers from severe sign problem away from the half filling. We therefore focus on the vicinity of the half filing case and study the doping effect in a Kondo insulator simply by tuning the energy levels of a selected number of local $f$-electron sites. For sufficiently depleted (or doubly occupied) local $f$-levels, this is equivalent to replace the local magnetic ions by the nonmagnetic conterparts. We find no severe sign problem in the considered temperature and tuning range. This allows us to study the evolution of the electronic states with arbitrary number of Kondo holes. We are able to reproduce the predicted bound states around the Kondo hole site obtained in previous mean-field calculations and find that the hybridized heavy electron states are destroyed in the neareast-neighbor sites. We further study the electronic transport with increasing doping and confirm the percolation transition from the Kondo insulator state in the dense Kondo lattice to the single-ion Kondo behavior in the diluted limit.

\section*{Results}

We start with the following modified Hamiltonian for the periodic Anderson lattice,
\begin{equation}
H = -t\sum_{<ij>,\sigma}(c_{i\sigma}^{\dagger}c_{j\sigma}+\text{H.c.})+V\sum_{i,\sigma}(c_{i\sigma}^{\dagger}f_{i\sigma}+\text{H.c.})+U\sum_{i}(n_{i\uparrow}^{f}-\frac12)(n_{i\downarrow}^{f}-\frac12)+\sum_{I,\sigma}{}^\prime\epsilon_f^If_{I\sigma}^{\dagger}f_{I\sigma},
\end{equation}
where $c_{i\sigma}^{\dagger}$ ($f_{i\sigma}^{\dagger}$) and $c_{i\sigma}$ ($f_{i\sigma}$) are the creation and annihilation operators for the conduction and $f$-electrons, respectively. $n_{i\sigma}^{c}= c_{i\sigma}^{\dagger} c_{i\sigma}$ and $n_{i\sigma}^f= f_{i\sigma}^{\dagger} f_{i\sigma}$ are the corresponding number operators for the spin-$\sigma$ component on the $i$-th site, $t$ is the hopping parameter of the conduction electrons between the nearest-neighbor sites, $V$ describes the local hybridization between the conduction and $f$-electrons, and $U$ is the local Coulomb interaction on the local $f$-site. $\epsilon_f^I$ represents the $f$-electron energy at the impurity sites $I$, which is tuned away from zero to resemble the effect of chemical substitution. The sum $\sum{}^\prime$ in the last term is only over the impurity sites. We have modified the DQMC code implemented in QUantum Electron Simulation Toolbox \cite{tomas2012advancing} for the tuning and carried out numerical simulations on a $12\times12$ square lattice with a discretization of the imaginary time, $\Delta\tau=0.25$. For $\epsilon_f^I=0$, the model reduces to a clean Kondo lattice, which has no sign problem due to the particle-hole symmetry. For large $|\epsilon_f^I|$, the local $f$-electron occupation is effectively empty or full, so the $f$-electron degree of freedom at the impurity site plays essentially no role in the simulation and we also find no severe sign problem. The magnetic and electronic properties of the the $f$-electrons can be obtained through the equal-time spin correlation function \cite{hirsch1989antiferromagnetism,vekic1995competition},
\begin{equation}
c_f(l_{x},l_{y})=\frac{1}{N}(-1)^{l_{x}+l_{y}}\langle(n^f_{i+l\uparrow}-n^f_{i+l\downarrow})(n^f_{i\uparrow}-n^f_{i\downarrow})\rangle,
\end{equation}
and the imaginary time Green's function,
\begin{equation}
G_{f;i\sigma}(\tau) =<T_{\tau}f_{i\sigma}(\tau)f_{i\sigma}^{\dagger}(0)>.
\end{equation}
The local density of states of the $f$-electrons, $\rho_{f;i\sigma} (\omega)$, can be obtained from $G_{f;i\sigma}(\tau)$ using the maximum entropy method,
\begin{equation}
G_{f;i\sigma}(\tau)=\int_{-\infty}^{\infty}d\omega\frac{e^{-\omega\tau}}{1+e^{-\beta\omega}} \rho_{f;i\sigma}(\omega).
\end{equation}
The same can be done for the conduction electrons. For simplicity, we set $t=1$ and $U/t=6$ in the numerical calculations and use the natural units with the Boltzmann constant $k_B=1$. The large value of $U$ is chosen for typical Kondo lattice systems \cite{jiang2014universal,shim2007modeling}.

\begin{figure}[t]
\centering
\includegraphics[width=\linewidth]{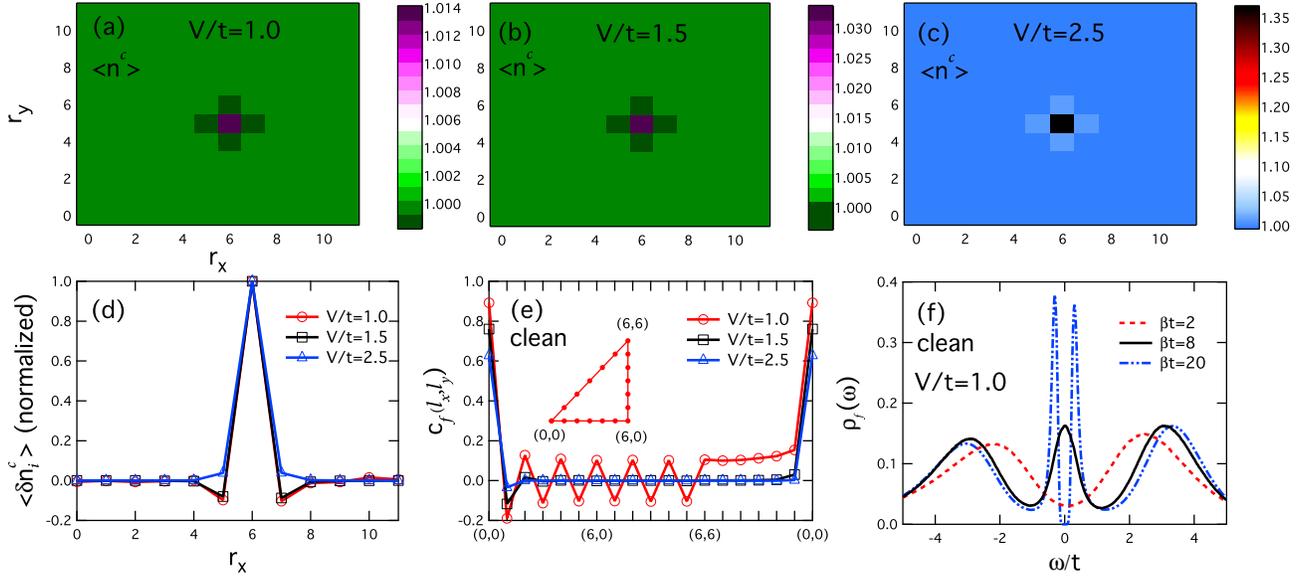}
\caption{Perturbation of a single Kondo hole to the charge density of the conduction electrons. (a-c) Contour plots of the occupation number of the conduction electrons, $\langle n^{c}\rangle$, with a single Kondo hole for $V/t=$1.0, 1.5 and 2.5, respectively. (d) The perturbation, $\langle n^{c}\rangle-1.0$, along $r_{x}$ across the Kondo hole site as shown in (a-c). (e) The spin correlation function $c(l_x,l_y)$ of the $f$-electrons in the clean lattice along a triangular path shown in the inset. (f) The local density of states of the $f$-electrons in the clean lattice for $V/t=1.0$ with varying temperature. Other parameters are $\beta t=20$ in (a-e) and $\epsilon_f^I/t=30$ in (a-d).}
\label{fig1}
\end{figure}

To provide a basis for comparison, we first present the results for the clean lattice ($\epsilon_f^I=0$) in Figs.~\ref{fig1}(e) and \ref{fig1}(f). The spin correlation function at low temperature ($T/t=0.05$) exhibits clear antiferromagnetic spatial oscillation for $V/t=1.0$, indicating antiferromagnetic long-range order in the finite lattice for weak hybridization. For stronger hybridization with $V/t=1.5$ and 2.5, the antiferromagnetic oscillation is suppressed and, as expected, the system shows a transition from the antiferromagnetically long-range ordered state to the paramagnetic (disordered) state with increasing hybridization due to the competition between the collective Kondo hybridization and the induced RKKY (Ruderman-Kittel-Kasuya-Yosida) exchange coupling among the $f$-moments \cite{doniach1977kondo}. Fig.~\ref{fig1}(f) plots the $f$-electron local density of states at $V/t=1.0$. We see for $T/t=0.5$ two broad Hubbard peaks located at $\omega=\pm U/2=\pm 3$. As temperature decreases, a narrow resonance peak first appears  at $T/t=0.125$, reflecting the onset of coherence, and then splits into two sharp peaks at $T/t=0.05$. The split of the resonance peak is a special feature of the Anderson lattice model and originates from the collective hybridization between the conduction band and the effective $f$-electron flat band near the Fermi energy. The emergence of the gap feature indicates that collective hybridization already takes place in the antiferromagnetic state.

\begin{figure}[t]
\centering
\resizebox{10cm}{!}{\includegraphics{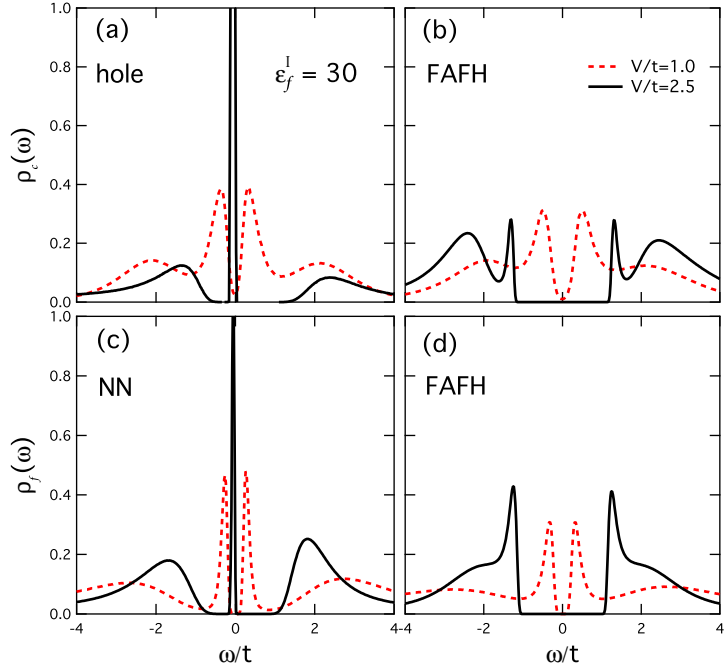}}
\caption{Bound states in the local densities of states around the Kondo hole. (a,b) The local density of states $\rho_{c}(\omega)$ of the conduction electrons at the Kondo hole site and far away from hole (FAFH) site. (c,d) The local density of states $\rho_{f}(\omega)$ of the $f$-electrons at the nearest-neighbor (NN) site and far away from hole (FAFH) site for different hybridization strengths $V/t=1.0$ and 2.5. Other parameters are $\epsilon_f^I/t=30$ and $\beta t=20$.
\label{fig2}}
\end{figure}

The effect of doping can be readily obtained by comparing the results for the clean Anderson lattice ($\epsilon_f^I=0$) and that with a finite $\epsilon_f^I$. Figs. \ref{fig1}(a-d) plot the spatial distribution of the average $\langle n_{i\sigma}^c\rangle$, the occupation number of the conduction electrons, for $\epsilon_f^I/t=30\gg D=4t=4$. This resembles the situation of a Kondo hole, as the local $f$-electron occupation number at the impurity site is almost zero. On the other hand, the local occupation of the conduction electrons, $\langle n_{I\sigma}^c\rangle$, is strongly enhanced and increases rapidly with increasing $V$. This may be understood if we integrate out the local $f$-electron degree of freedom at the impurity site. We obtain effectively a local attractive potential, $\delta\epsilon_c^I \propto -V^2/|\epsilon_f^I|$, which tends to trap the conduction electrons on the Kondo hole site, causing the increase of $\langle n_{I\sigma}^c\rangle$. For small $V/t=1.0$ and 1.5, we see that $\langle n_{i\sigma}^c\rangle$ is enhanced at the impurity site while slightly reduced in the nearest-neighbor sites. The charge density oscillation is also associated with this potential scattering, reflecting the effect of Friedel oscillation of free conduction electrons. This is possible because for sufficiently small $V$, the scattering potential is relatively stronger and overcomes the hybridization energy, $\Delta_h\propto D e^{-aUt/V^2}$, where $a$ is a constant of the order of unity. As expected, there is no oscillation for strong hybridization at $V/t$=2.5. $\langle n_{i\sigma}^c\rangle$ exhibits a rapid decay with increasing distance away from the Kondo hole. We note that unlike previous mean-field studies \cite{figgins2011defects,toldin2013disorder,zhu2012local}, we observe no evident spatial oscillation in the hybridization function \cite{hamidian2011kondo}, indicating that the perturbation is suppressed in our model, possibly due to the half filling.

\begin{figure}[t]
\centering
\resizebox{10cm}{!}{\includegraphics{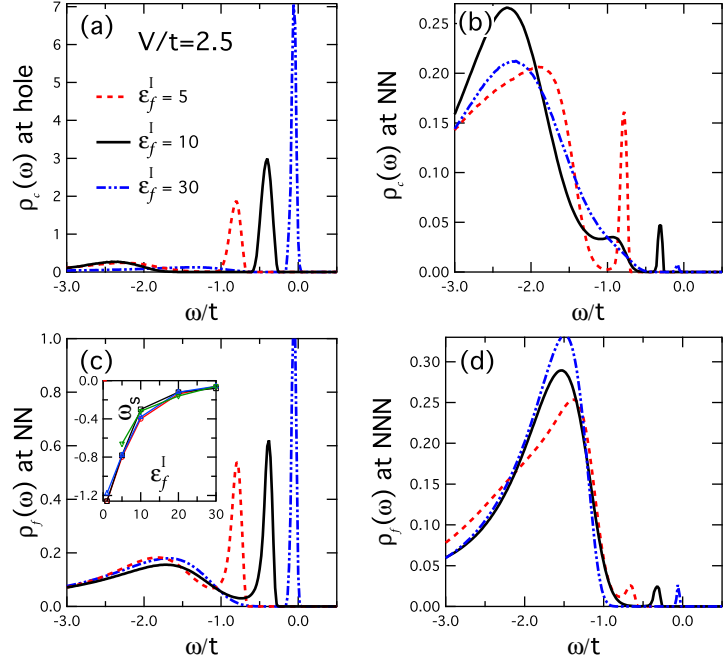}}
\caption{The variation of the bound states with $\epsilon_f^I$. (a,b) The local density of states $\rho_{c}(\omega)$ of the conduction electrons at the impurity site and the nearest-neighbor (NN) site. (c,d) The local density of states, $\rho_{f}(\omega)$, of the $f$-electrons at the nearest-neighbor (NN) site and the next-nearest-neighbor (NNN) site. The impurity level is taken to be $\epsilon_f^I/t=5$, 10, 30. The inset in (c) shows the location $\omega_{s}$ of the bound state as a function of $\epsilon_f^I$. Other parameters are $V/t=2.5$ and $\beta t=20$.
\label{fig3}}
\end{figure}

The perturbation to the local electronic densities of states is presented in Fig.~\ref{fig2} and compared with that far away from the Kondo hole site (FAFH) for $V/t=1.0$ and 2.5. The latter is the same as that in the clean lattice. As shown in Figs.~\ref{fig2}(a) and \ref{fig2}(c), a sharp peak emerges inside the hybridization gap for both $\rho_c$ at the Kondo hole site and $\rho_f$ at the nearest-neighbor sites for $V/t=2.5$, in contrast to the unperturbed results far away from the Kondo hole site. The resonance peak gives the well-known bound states and is missing if hybridization strength at the hole site is set to zero (no impurity scattering). For small $V$, the bound states are absent due to weak impurity scattering as discussed above and the peaks at the edges of the hybridization gap reflect the hybridization between the conduction electrons and the $f$-moments, similar to those in the clean lattice. The suppression of these hybridization states at the gap edges for $V/t=2.5$ is a manifestation of the destruction of the heavy electrons due to the presence of the nearby Kondo hole. This suggests that the heavy electron emergence is not a single-site property but a collective phenomenon that involves correlations among neighboring sites in the dense Kondo lattice.

Fig. \ref{fig3} further compares the densities of states at different sites and for varying $\epsilon_f^I$. We see that the height of the resonance peak is reduced substantially by over an order of magnitude further away from the impurity site. Nevertherless, as $\epsilon_f^I$ increases, the bound states move uniformly towards the center of the hybridization gap and become more and more pronounced. The change in their location is shown as a function of $\epsilon_f^I$ in the inset of Fig.~\ref{fig3}(c). We obtain similar curves for the peaks in both $\rho_c$ at the hole site and $\rho_f$ at the NN sites, indicating their common origin. As expected, the bound state approaches $\omega=0$ as $\epsilon_f^I\rightarrow \infty$, consistent with the mean-field prediction.

\begin{figure}[t]
\centering
\resizebox{10cm}{!}{\includegraphics{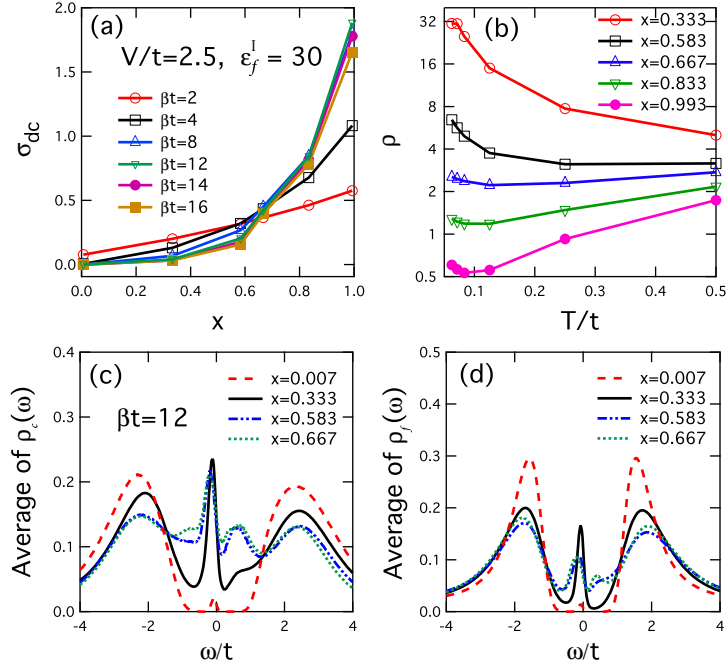}}
\caption{Percolation transition from the Kondo insulator state to the single-ion Kondo state with increasing doping. (a) The $dc$ conductivity $\sigma_{dc}$ as a function of doping $x$ at different temperatures. (b) The resistivity $\rho=1/\sigma_{dc}$ as a function of temperature for various doping $x$. (c,d) The average $\rho_{c}(\omega)$ and $\rho_{f}(\omega)$ at $\beta t=12$ for different $x$. Other parameters are $V/t=2.5$ and $\epsilon_f^I/t=30$.
\label{fig4}}
\end{figure}

Next, we study how increasing doping may change the electronic and transport properties by tuning the system from the Kondo insulator to the diluted limit. We add more impurity sites whose positions are chosen randomly in the numerical simulation and calculate the $dc$ conductivity of the doped Anderson lattice using the approximation,
\begin{equation}
\sigma_{dc}=\frac{\beta^2}{\pi}g_{xx}(\tau =\beta /2),
\end{equation}
in which $g_{xx}(\tau)$ is the current-current correlation function in imaginary time,
\begin{equation}
g_{xx}(\tau)= -\langle T_{\tau}j_{x}(\tau)j_{x}(0) \rangle,
\end{equation}
with
\begin{equation}
j_{x}(\tau)=it\sum_{l\sigma}\left[c_{l+x,\sigma}^{\dagger}(\tau)c_{l,\sigma}(\tau)-c_{l,\sigma}^{\dagger}(\tau)c_{l+x,\sigma}(\tau)\right].
\end{equation}
The resulting conductivity, $\sigma_{dc}$, is plotted in Fig.~\ref{fig4} as a function of doping $x$ for different temperatures. The data presented here are an average over only 5 samples of random configurations. We have checked the results up to 20 samples and found a numerical error of less than 5\%. We see all  curves cross at a critical doping, $x_c\approx 0.6$, indicating opposite temperature dependence below and above $x_c$. Moreover, our analysis of finite size effect shows no significant change in $x_c$ with increasing $L$ (see Supplementary Fig. S1). For $x < 0.6$,  $\sigma_{dc}$ is small and the resistivity, $\rho=1/\sigma_{dc}$, increases as temperature decreases, so that the system is in the Kondo insulator phase. As $x$ becomes larger than 0.6, $\sigma_{dc}$ increases rapidly with $x$ as the system approaches the single-ion Kondo limit. In this regime, $\rho(T)$ first decreases with deceasing temperature and exhibits metallic behavior at high temperatures down to about $T<0.1$, below which the resistivity becomes insulating like, with a minimum signalling the typical single-ion Kondo behavior in the diluted limit. Our calculations therefore cover the whole doping range from the dense Kondo lattice to the single-ion Kondo limit.

To understand the connection between the transport property and the bound states, we calculate the average densities of states of both the conduction and $f$-electrons on all sites. The results are plotted in Figs.~\ref{fig4}(c) and \ref{fig4}(d). We see that the average impurity states exhibit broad peaks inside the hybridization gap, similar to the mean-field results. For $x=0.333$, 0.583 and 0.667, the gap is gradually filled in, yielding a finite density of states at $\omega=0$, in contrast to the insulating like behavior shown in our calculated resistivity curve. Hence the average density of states cannot be used directly to explain the transport properties. This points to the percolative nature of the electron transport. In the theory of site percolation \cite{aharony2003introduction}, adjacent bound states could form a cluster. When the size of the cluster grows and eventually percolates through the whole lattice after stochastic doping of enough sites, the system reaches a percolation threshold where a phase transition takes place. For the square lattice of infinite size in two dimension with only nearest-neighbor hopping, the percolation threshold is $x_c\approx0.593$ \cite{aharony2003introduction}. Increasing the number of coordination or dimension can reduce the threshold \cite{schlottmann1996metal}. In our two dimensional $12\times 12$ system, bound states are only pronounced at the hole site for the conduction electrons. Since only nearest-neighbor hopping is allowed, the impurity sites have to be connected to form a network in order for the conduction electrons to move from one side to the other side of the lattice. As shown in Fig.~\ref{fig4}, the critical doping $x_c\approx 0.6$ obtained in our resistivity calculations is in good agreement with the percolation threshold $x_c$. Moreover, previous theories have also predicted for $T=0$ and $x>x_c$ a power law scaling, $\sigma_{dc}\propto (x-x_c)^{\mu}$, with $\mu\approx1.3$ \cite{clerc1990electrical}, consistent with our fit giving $\mu\approx 1.44$ for $L=12$ and $\beta t=16$ (see Supplementary Fig. S2). The doping-induced insulator-metal transition is therefore an indication of site percolation of the conducting electrons through the bound states, which governs the charge transport in the heavily doped Kondo lattice. Our results confirm the previous theoretical and experimental speculations. Along the percolation path, collective hybridizations are destroyed around the impurity sites, providing an explanation to the crossover from the heavy electron physics to the single-ion Kondo physics.

\section*{Conclusion}
We use the determinant Quantum Monte Carlo (DQMC) method to study the doping effects in the two-dimensional periodic Anderson lattice. We observe the suppression of the hybridization feature in the vicinity of the Kondo hole, which suggests a physical mechanism to destroy the heavy electron states with increasing doping. Our results also confirm the bound states appearing around the Kondo hole, but show no obvious signature of the hybridization oscillation predicted in the mean-field calculations. We further study the evolution of the electron transport with increasing doping and demonstrate the percolative nature of the insulator-metal transition from the dense Kondo lattice to the single-ion Kondo limit. We obtain a critical doping that is consistent with the predicted threshold for site percolation.


\section*{Acknowledgements}
This work was supported by the National Natural Science Foundation of China (NSFC Grant No. 11522435), the State Key Development Program for Basic Research of China (Grant No. 2015CB921303), the Strategic Priority Research Program (B) of the Chinese Academy of Sciences (Grant No. XDB07020200) and the Youth Innovation Promotion Association CAS.

\section*{Author contributions statement}
Y.Y. conceived the idea. Y.Y. and L.W. performed the research and wrote the paper.

\section*{Additional information}

\textbf{Competing financial interests:} The authors declare no competing financial interests.

\clearpage
\section*{Supplementary Materials}
\renewcommand\thefigure{S\arabic{figure}} 
\setcounter{figure}{0}

Here we provide further analysis on the finite size effect and the critical scaling behavior of the $dc$ conductivity. \\

\begin{figure}[ht]
\centering
\resizebox{9.5cm}{!}{\includegraphics{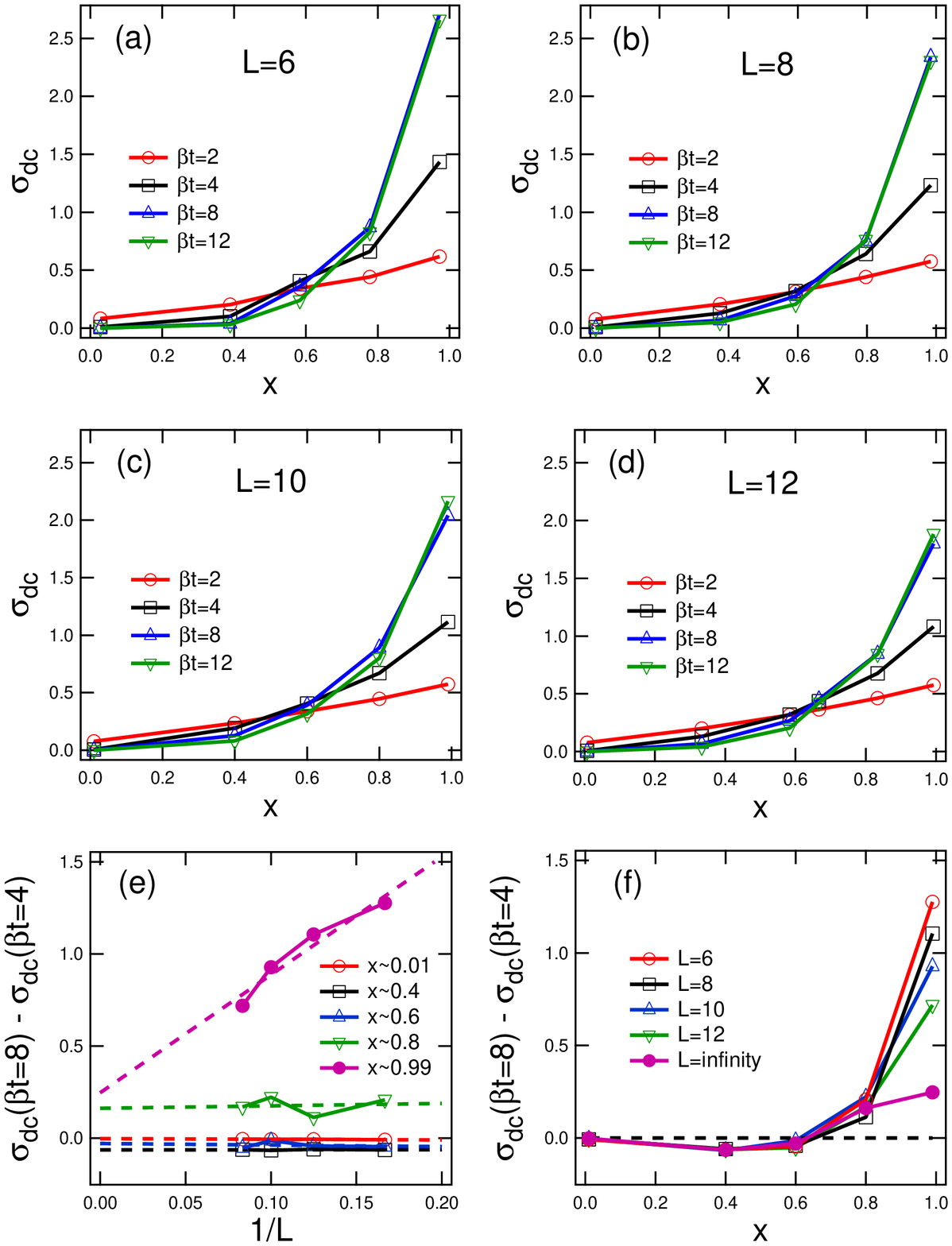}}
\caption{Lattice size dependence of the $dc$ conductivity $\sigma_{dc}$. (a-d) $\sigma_{dc}$ as a function of doping $x$ at different temperatures for the lattice size $L=6$, 8, 10, and 12. (e) $\sigma_{dc}(\beta t=8)-\sigma_{dc}(\beta t=4)$ as a function of $L$ for different dopings, showing the size dependence of the metallic or insulating behavior. The dashed lines illustrate linear extropolations to the numerical data. (f) Doping dependence of $\sigma_{dc}(\beta t=8)-\sigma_{dc}(\beta t=4)$ for different lattice sizes, showing slight change in the critical doping $x_c\approx 0.6$ for the insulator-to-metal transition.}
\label{figS1}
\end{figure}

Figure \ref{figS1} compares the difference in the $dc$ conductivity, $\sigma_{dc}(\beta t=8)-\sigma_{dc}(\beta t=4)$, for $L=6, 8, 10, 12$. We see that the critical doping is almost unchanged with increasing lattice size $L$. We note that the extrapolation at $x=0.99$ cannot be trusted as the doping is actually $(L^2-1)/L^2$, which varies with $L$. Similar variations exist for other doping, but are less crucial.

\newpage

\begin{figure}[ht]
\centering
\resizebox{9cm}{!}{\includegraphics{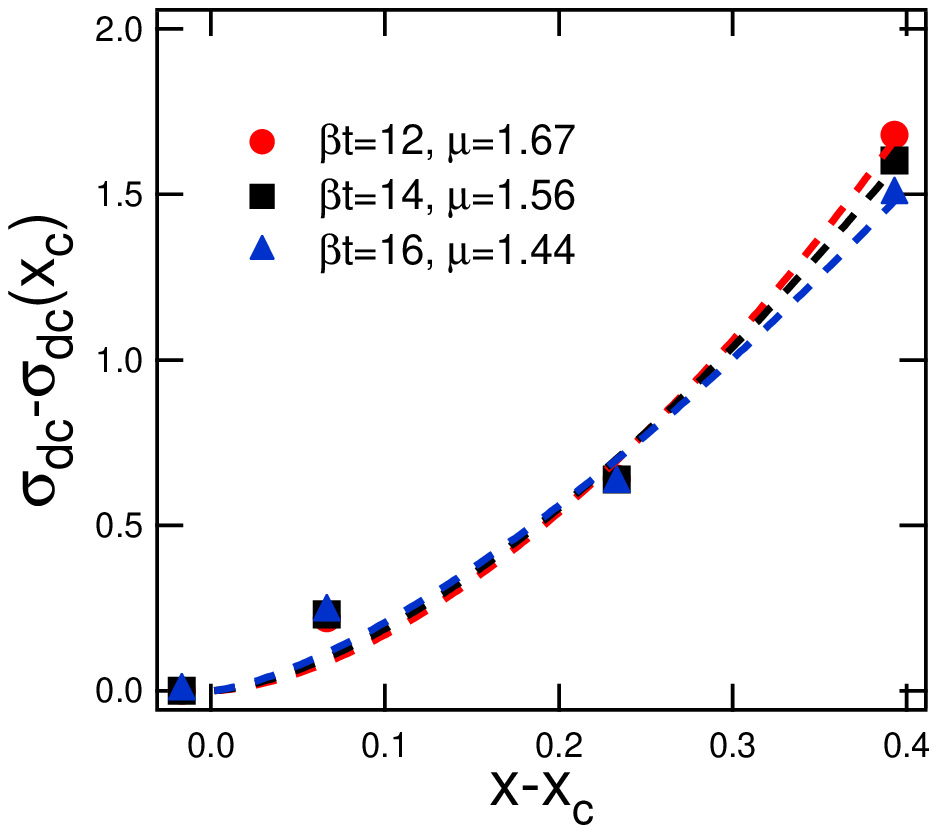}}
\caption{Fit to the $dc$ conductivity, $\sigma_{dc}-\sigma_{dc}(x_c)$, with $(x-x_c)^{\mu}$ for $L=12$ and $\beta t=12$, 14 and 16.
\label{figS2}}
\end{figure}

Figure \ref{figS2} shows a power law fit to the $dc$ conductivity, $\sigma_{dc}-\sigma_{dc}(x_c)$, with $(x-x_c)^{\mu}$ for $L=12$ and $\beta t= 12$, 14 and 16. The fit yields $\mu=1.67$, 1.56 and 1.44, respectively, while previous theories have predicted $\mu\approx 1.3$ for zero temperature and 2D infinite lattice. Our results are close to the predicted values.

\end{document}